\documentclass[letter]{aa}

\usepackage{graphicx}
\usepackage{txfonts}
\usepackage{natbib}

\begin{document} 

\title{Atomic oxygen abundance toward Sagittarius B2\thanks{We dedicate this manuscript to Tom Phillips, who was an inspiration to all of us. Tables with the data used for Figures 2 and B.3 are available at the CDS via anonymous ftp to cdsarc.u-strasbg.fr (130.79.128.5) or via http://cdsarc.u-strasbg.fr/viz-bin/cat/J/A+A/XXX/XX
  }}

\author{Dariusz C. Lis\inst{1}, Paul F. Goldsmith\inst{1}, Rolf G\"{u}sten\inst{2}, Peter Schilke\inst{3}, Helmut Wiesemeyer\inst{2}, Youngmin Seo\inst{1}, and Michael~W.~Werner\inst{1}}

\institute{Jet Propulsion Laboratory, California Institute of Technology, 4800 Oak Drove Drive, Pasadena, CA 91109, USA
  \and Max-Planck-Institut für Radioastronomie, Auf dem H\"{u}gel 69, D-53121 Bonn, Germany
  \and I. Physikalisches Institut, Universität zu K\"{o}ln, Z\"{u}lpicher Stra\ss e 77, D-50937 Köln, Germany}

\date{Received 11 November 2022; accepted 23 December 2022.}

\abstract{
  A substantial fraction of oxygen in diffuse clouds is unaccounted for by observations and is postulated to be in an unknown refractory form, referred to as unidentified depleted oxygen (UDO), which, depending on the local gas density, may contribute up to 50\% of the total oxygen content. Previous Infrared Space Observatory (ISO) observations suggest that a significant fraction of oxygen in even denser, translucent clouds may be in atomic form.
  We have analyzed velocity-resolved archival SOFIA observations of the 63~$\mu$m fine-structure [O\,{\sc i}] transition toward the high-mass star-forming region Sgr B2(M) in the Central Molecular Zone. The foreground spiral-arm clouds as well as the extended Sgr B2 envelope between the Sun and the background dust continuum source produce multiple [O\,{\sc i}] absorption components, spectrally separated in velocity space. The gas-phase atomic oxygen column density in foreground clouds toward Sgr B2 is well correlated with the total hydrogen column density, with an average atomic oxygen abundance of $(2.51 \pm 0.69) \times 10^{-4}$ with respect to hydrogen nuclei. This value is in good agreement with the earlier ISO measurements on the same line of sight, and is about 35\% lower than the total interstellar medium oxygen abundance in the low-density warm gas, as measured in the UV. We find no evidence that a significant fraction of the oxygen on the line of sight toward Sagittarius B2 is in the form of UDO.
}
  
\keywords{Astrochemistry -- ISM: abundances -- ISM: atoms -- ISM: clouds -- ISM: lines and bands -- ISM: molecules}

\titlerunning{Atomic oxygen}
\authorrunning{Lis et al.}

\maketitle

\section{Introduction}

\label{sec:intro}

A long-standing problem in our understanding of the quiescent dense interstellar medium (ISM) has been the difficulty of accounting for the gas-phase abundances of carbon and oxygen. Since the first calculations of ion-molecule reaction schemes \citep{herbst73, dalgarno76}, there have been theoretical predictions indicating that the fundamental reservoirs of these elements are the molecular species CO, O$_2$, and H$_2$O. The local gas-phase oxygen abundance is assumed to be about twice that of carbon. This comes from the observed stellar values, modified by the depletion seen in local ISM diffuse clouds, such as those toward $\zeta$ Ophiuchi and HD 154368 \citep{snow96a, snow96b, cardelli93}, that is, the carbon abundance [C] = $1.66 \times 10^{-4}$ and $1.32 \times  10^{-4}$, respectively, and the oxygen abundance [O] = $2.88 \times 10^{-4}$, which gives an average [C]/[O] = 0.51.
So, nearly all carbon should be in CO, with plenty of oxygen left over for O$_2$, and H$_2$O.
  
Millimeter-wave measurements have indeed confirmed the large abundance of CO at about $10^{-4}$ of H$_2$ (e.g., \citealt{lada94}). However, extensive \emph{Herschel} observations \citep{vandishoeck21} have shown that the H$_2$O abundance is universally low. Even in warm outflows and shocks, the water abundance is only $\sim 10^{-6}$ with respect to H$_2$, much less than the expected value of $4 \times 10^{-4}$  if all volatile oxygen is in water. Only in very hot gas (> 1000 K) have water abundances close to $10^{-4}$ been derived. For O$_2$, low gas-phase abundances have been suggested by early Submillimeter Wave Astronomy Satellite (SWAS)
and Odin observations \citep{goldsmith00,larsson07}. \emph{Herschel} pushed the limits even lower, with O$_2$ detections reported only in the Orion shock \citep{goldsmith11,chen14} and in the $\rho$ Ophiuchi cloud \citep{liseau12,larsson17}, where an abundance of $5 \times 10^{-8}$ with respect to H$_2$ has been derived. In other sources, limits as low as $6 \times 10^{-9}$ ($3 \sigma$) were reported (NGC 1333 IRAS4A; \citealt{yildiz13}).
  
A combined analysis of water vapor, water ice, and O$_2$ limits in cold clouds thus indicates that a large fraction of the oxygen is unaccounted for. A number of possible explanations have been hypothesized \citep{vandishoeck21}. Within the simple water chemistry models, the only solution is to have a short pre-stellar stage of only 0.1~Myr to prevent all oxygen from being turned into water. An alternative is for dense cores to have a small fraction of large grains ($>1~\mu$m), which prevents more than 50\% of the water ice from being observed through infrared absorption spectroscopy. However, this solution does not apply to hot cores and shocks, where the large icy grains should have sublimated and where a large fraction of oxygen is also missing. Another option is, therefore, that oxygen is in some refractory form called unidentified depleted oxygen (UDO), which consists of material that does not vaporize or atomize even in strong shocks (up to 1000~K).

A schematic overview of the oxygen budget in diffuse and translucent clouds is shown in Fig.  3 of \cite{whittet10}. In diffuse clouds, most of the oxygen is in atomic form, with silicates and oxides contributing up to about 100 pm (out of a total of 575 ppm relative to H nuclei). The UDO wedge starts at hydrogen nucleus densities of about 0.1 cm$^{-3}$ and increases to about 150 ppm ($\sim 25\%$ of the total oxygen) at densities of about 7 cm$^{-3}$, corresponding to the effective observational limit on depletion studies in the UV. In higher-density clouds ($n_{\rm H} \sim 1000~{\rm cm}^{-3}$), gas-phase CO is expected to contribute at about 50 ppm, with ices, and silicates and oxides
contributing at about 100 ppm each. The atomic oxygen contribution is predicted to be small, leaving about 300 ppm in the form of UDO ($\sim 50\%$ of the total oxygen). This denser gas is not accessible in the UV due to extinction, but can be studied using far-infrared spectroscopy.

Atomic oxygen is an important part of the oxygen budget, and a number of studies aimed at constraining its abundance have been carried out. The [O\,{\sc i}] 63~$\mu$m fine-structure line was observed in the diffuse ISM using the Infrared Space Observatory (ISO)  Long Wavelength Spectrometer (LWS)
by means of absorption spectroscopy against bright background dust continuum sources. Such studies were limited by the relatively low spectral resolution of the LWS instrument (10 km\,s$^{-1}$ in the Fabry-Perot mode with the maximum entropy 
deconvolution).
However, in some sources, the foreground absorption is well separated in velocity space from the background source, allowing a determination of the oxygen abundance in the foreground clouds \citep{vastel00, lis01, vastel02}.

The [O{\,\sc i}] 63~$\mu$m fine-structure line emission has been widely used as a tracer of star formation both in Galactic
sources \citep{liseau99, oberst11, karska14} and in external galaxies \citep{malhotra01, dale04, gonzalez12, farrah13}. Comparison of the 63~$\mu$m and 145~$\mu$m line intensities \citep{stacey83} and 63~$\mu$m studies with higher spectral resolution \citep{kramer98, boreiko96, leurini15, schneider18, mookerjea19} suggest that the lower-lying 63~$\mu$m line observed in emission is optically thick.
\cite{goldsmith21} find that approximately half of the 12 sources observed with the German REceiver for Astronomy at Terahertz Frequencies (GREAT)
instrument on the Stratospheric Observatory for Infrared Astronomy (SOFIA)
showed clear evidence of self-absorption profiles,
indicating the presence of large column densities of low-excitation atomic oxygen with $\rm N(O^0_{le}) = 2 - 7 \times 10^{18}~cm^{-2}$.
Much of this is in regions that would typically be assumed to be totally molecular, but which in fact have $X({\rm O^o}) \simeq 10^{-5}$. The low-excitation foreground gas can be studied by means of absorption spectroscopy toward bright background dust continuum sources.


Sgr B2 is one of the brightest far-infrared continuum sources in the Galaxy, and thus an excellent target for absorption studies. The differential rotation of the Milky Way allows spectral features from gas clouds at different galactocentric radii to be separated in velocity
(\citealt{greaves94}; Fig. 4).  Even at the limited spectral resolution of the ISO LWS, the Sgr B2(M) [O\,{\sc i}] spectrum could be decomposed into three foreground velocity components (Fig. 3 of \citealt{lis01}), and the atomic oxygen column density was shown to be correlated with the CO column density, as expected if the two species are well mixed spatially.  An average atomic oxygen abundance of $2.7 \times 10^{-4}$ with respect to hydrogen nuclei was derived in the molecular phase \citep{lis01}. 

The ISO study was limited by the spectral resolution of the LWS instrument, which resulted in blending of multiple velocity components. The GREAT instrument on SOFIA offers tremendous improvements in sensitivity and spectral resolution at 63~$\mu$m over ISO LWS (see, e.g., \citealt{wiesemeyer16,goldsmith21} for velocity-resolved [O\,{\sc i}] observations of other sources).  In the present paper, we revisit the atomic oxygen abundance in the foreground clouds on the sightline toward Sgr B2 using archival SOFIA observations. The high spectral resolution of GREAT allows, for the first time, the [O\,{\sc i}] emission from individual line-of-sight clouds to be separated, velocity intervals affected by saturated absorption to be correctly masked, and accurate atomic oxygen column densities and abundances to be derived.

\section{Observations}
\label{sec:obs}

We used publicly available SOFIA/GREAT \citep{heyminck12} observations of the 63~$\mu$m fine-structure [O\,{\sc i}] line from the NASA Infrared Processing and Analysis Center (IPAC)
SOFIA Science Archive\footnote{https://irsa.ipac.caltech.edu/Missions/sofia.html; AOR ID 03\_0088.}. The data downloaded from the archive were re-reduced using the latest version of the GREAT pipeline (see Appendix A). The [O\,{\sc i}] spectrum is centered at the position of Sgr B2M, $\rm 17^h 47^m 20.16^s; -28^d 23^\prime 04.5^{\prime\prime}$ (J2000).

\begin{figure}
   \centering
   \includegraphics[trim=2cm 2.5cm 3cm 4cm, clip=true, width=0.99\columnwidth]{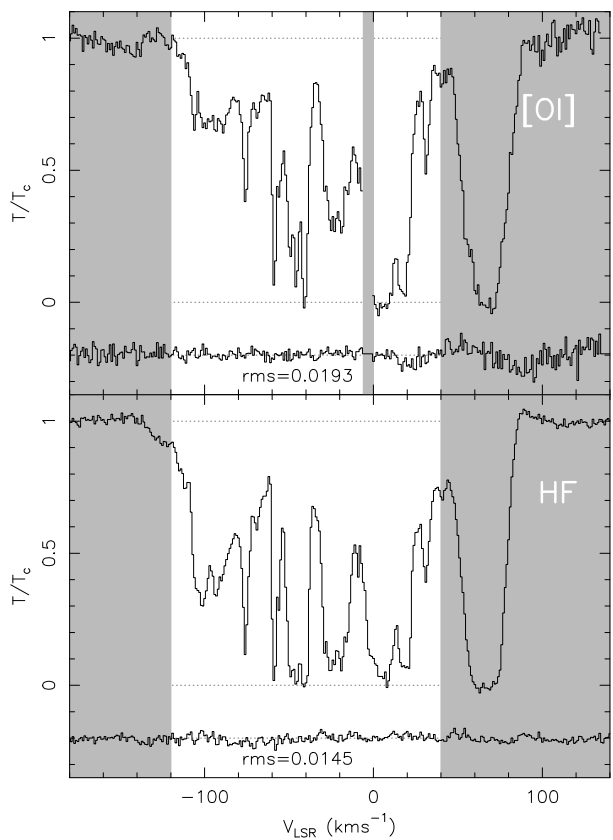}
   \caption{
     Spectra of [O\,{\sc i}] and HF absorption toward Sgr B2(M) divided by the corresponding continua. (Upper) SOFIA/GREAT [O\,{\sc i}] spectrum. The lower histogram shows the difference between two independent data subsets divided by 2, which is a measure of the uncertainty in the [O\,{\sc i}] spectrum. Gray areas show velocities excluded from the analysis. This includes the Sgr B2 envelope at velocities greater than 40 km\,s$^{-1}$ and the region between -6 and 0 km\,s$^{-1}$, where the [O\,{\sc i}] spectrum is contaminated by telluric absorption. (Lower) Average \emph{Herschel}/HIFI spectrum of HF. The lower histogram shows the difference between the two independent observations divided by 2, which is a measure of the uncertainty in the average spectrum. } \label{fig1}
\end{figure}

Figure~\ref{fig1} (upper panel) shows the final [O\,{\sc i}] 63~$\mu$m spectrum divided by the continuum, resampled to 1~km\,s$^{-1}$ spectral resolution. Local standard of rest (LSR) velocities between $-120$ and +40~km\,s$^{-1}$ correspond to the foreground gas, while those greater than 40~km\,s$^{-1}$ correspond to the envelope of the Sgr~B2 cloud. The higher gas densities present in this component make the excitation and the resulting column densities uncertain. Consequently, we excluded these velocities from the analysis. Velocities between $-6$ and 0~km\,s$^{-1}$, where the [O\,{\sc i}] spectrum is contaminated by telluric absorption, were also excluded. To estimate the noise level in the [O\,{\sc i}] absorption region, we split the data into two independent subsets with comparable integration times. The lower histogram shows the difference spectrum between the two subsets divided by 2, which is a measure of the uncertainty in the final [O\,{\sc i}] spectrum. The difference spectrum is flat over most velocities. The rms computed in the $-130$ to 40 km\,s$^{-1}$ velocity range is 0.0193, and we used this value as the uncertainty of the [O\,{\sc i}] line-to-continuum ratio in the analysis. The rms increases toward the edges, where the two local oscillator (LO)
settings used in the observations do not fully overlap, resulting in an effectively shorter integration time. In addition, a higher rms is seen in the difference spectrum at positive velocities, where the foreground absorption may be contaminated by wings of the [O\,{\sc i}] emission from the Sgr~B2 envelope. The full width at half maximum (FWHM) SOFIA beam size at the [O\,{\sc i}] frequency is $\sim$ 6.6$^{\prime\prime}$.

We used archival \emph{Herschel} Heterodyne Instrument for the Far-Infrared (HIFI)
\citep{degraauw10}  observations of hydrogen fluoride (HF) to determine H$_2$ column densities in the various velocity components on the line of sight toward Sgr B2. Two independent data sets were used in the analysis, which allows an accurate quantification of the instrumental uncertainties. Both data sets, reduced using the latest HIFI instrument pipeline, were downloaded from the European Space Agency
\emph{Herschel} Science Archive\footnote{https://archives.esac.esa.int/hsa/whsa/} and imported into the Institut de radioastronomie millimétrique (IRAM) 
Gildas\footnote{https://www.iram.fr/IRAMFR/GILDAS/} software package for subsequent analysis. The first observation is a Band 5A spectral scan of Sgr B2(M), centered at the same position as the O\,{\sc i} spectrum (OBSID 1342204739). The second observation is a 4$^{\prime}$ long north-south strip taken in the double-beam-switch (DBS) observing mode (OBSID 1342205881). The DBS reference beams lie approximately $3^\prime$ to the east and west (i.e., perpendicular to the roughly north–south elongation of Sgr B2). Two spectra in the strip closest to Sgr B2(M) ($< 10^{\prime\prime}$ offsets) were averaged with uniform weighting to produce the final spectrum used in the analysis. We used spectra taken with the HIFI wide band spectrometer, which provided a spectral resolution of 1.1 MHz over a 4 GHz intermediate frequency (IF)
bandwidth.  The FWHM HIFI beam size at the HF frequency is $\sim$ 18$^{\prime\prime}$, about three times larger than the SOFIA beam size at the [O\,{\sc i}] frequency. However, the foreground clouds are expected to be extended on such angular scales and to fully cover the background continuum source. This conclusion is supported by the good agreement between the two independent HF spectra taken at positions offset by about half of the HIFI beam.

Figure~\ref{fig1} (lower panel) shows the final HF spectrum, an equally weighted average of the two instrumental polarizations and the two independent observations, resampled to a 1~km\,s$^{-1}$ velocity resolution. The lower histogram shows the difference between the spectral scan and DBS observations divided by 2, which is a measure of the uncertainty in the final HF spectrum. The difference spectrum is very flat and shows no residuals, even at positive velocities, where the background absorption may potentially be contaminated by the Sgr~B2 envelope emission (see the HF emission wing at velocities greater than 85 km\,s$^{-1}$). The rms computed in the $-130$ to 40 km\,s$^{-1}$ velocity range is 0.0145, and we used this value as the uncertainty of the HF line-to-continuum ratio in the analysis. 

\section{Results}
\label{sec:results}

To derive the oxygen optical depths and the corresponding column densities in the individual channels, we followed established procedures commonly used in the analysis of HIFI observations of light hydrides (e.g., \citealt{neufeld10, lis10, monje11}). We first derived the optical depth of the [O\,{\sc i}] and HF lines ($\tau = - ln[1 - T_L/T_C]$, where $T_L/T_C$ is the line-to-continuum ratio), assuming that the foreground absorption completely covers the continuum source and that all oxygen atoms are in the ground state. The spiral arm clouds on the line of sight toward Sgr B2 have moderate densities, up to a few times $10^4$ cm$^{-3}$ \citep{greaves94}.  This is lower than the critical density for the excitation of the 63~$\mu$m O\,{\sc i} line ($5.0 \times 10^5$~cm$^{-3}$ for collisions with H$_2$ and $7.8 \times 10^5$~cm$^{-3}$ for collisions with H; \citealt{lique18,goldsmith19});  the assumption that the entire population is in the ground state is thus well justified. Figure~\ref{figb1} shows the [O\,{\sc i}] optical depths and the corresponding column density as a function of velocity.

To determine the oxygen abundance, H$_2$ and H column densities in the line-of-sight clouds are required. Because of its unique thermochemistry, HF has been shown to be an excellent tracer of H$_2$ (e.g., \citealt{phillips10,neufeld10,sonnentrucker10,monje11}). The HF abundance with respect to H$_2$ in diffuse or translucent
clouds, determined from a comparison with CH observations, is in the range $(1.1 - 1.6) \times 10^{-8}$, with an average of $(1.4 \pm 0.17) \times 10^{-8}$ (multiple velocity components toward W51, W49N, and NGC6334I; \citealt{sonnentrucker10,emprechtinger12}).
We used this value to convert the HF column density to the H$_2$ column density (Fig.~\ref{figb2}). To characterize the atomic gas component on the line of sight toward Sgr B2, we used the H\,{\sc i} column densities of \cite{winkel17}. Figure~\ref{fig2} shows the total hydrogen nucleus column density as a function of velocity, along with the molecular and atomic contributions.

\begin{figure}
   \centering
   \includegraphics[trim=2cm 3cm 4cm 2cm, clip=true, width=0.99\columnwidth]{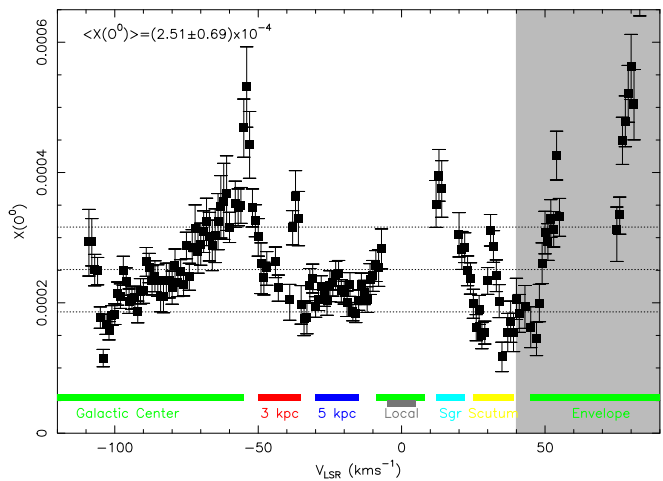}
   \caption{Atomic oxygen abundance relative to hydrogen nuclei as a function of velocity. The mean value of the 120 individual channels within the $-120$ and 40 km\,s$^{-1}$ velocity range is $2.51 \times 10^{-4}$, and the dispersion of the individual channels is $0.65 \times 10^{-4}$. The horizontal dotted lines mark the mean value and $\pm 1 \sigma$ dispersion computed from the ensemble of individual measurements. The black vertical error bars mark $\pm 1 \sigma$ uncertainties in the individual channels, as described in the text. Color bars mark velocity ranges corresponding to the 3 kpc, 5 kpc, Sagittarius, and Scutum arms (red, magenta, blue, and yellow, respectively). Velocities corresponding to the Galactic center gas are marked in green and those of the local gas in gray. Channels with saturated absorption are masked. An electronic table with the data used for this figure is available at the CDS.}
         \label{fig3}
\end{figure}

Figure~\ref{fig3} shows the gas-phase atomic oxygen abundance as a function of velocity, derived from the observations of the 63~$\mu$m line. The average abundance with respect to hydrogen nuclei, computed over the $-120$ to +40 km\,s$^{-1}$ velocity range, is  $(2.51 \pm 0.69) \times 10^{-4}$. The $\pm 1 \sigma$ dispersion of the individual measurements computed from the ensemble of 120 independent velocity channels is shown.
Black error bars are the formal $1 \sigma$ uncertainties of the individual measurements, computed by combining in quadrature corresponding uncertainties in the column densities of the atomic oxygen, atomic, and molecular hydrogen. They are typically smaller than the ensemble dispersion, suggesting the presence of variations in the local atomic oxygen abundance among different velocity components.

\section{Discussion}
\label{sec:discussion}

\begin{figure}
   \centering
   \includegraphics[trim=1.5cm 3.5cm 3.5cm 1.5cm, clip=true, width=0.99\columnwidth]{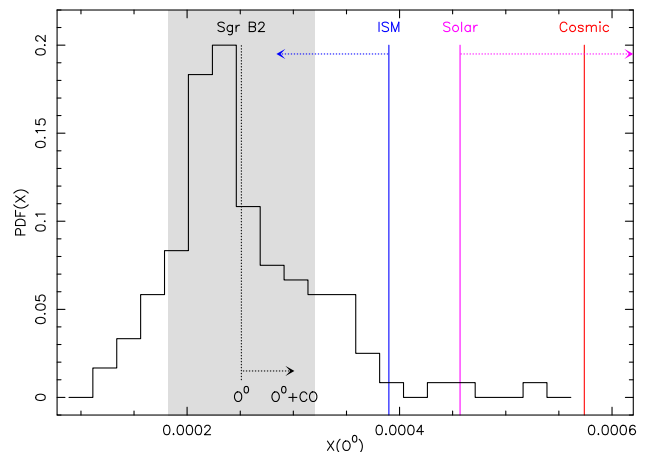}
   \caption{Normalized probability density function (PDF)
     of the O$^0$ abundances in 120 individual velocity channels toward Sgr~B2. The vertical dashed black line shows the mean abundance, the gray shaded area $\pm 1 \sigma$ departures from the mean, and the black arrow the corresponding approximate gas-phase oxygen content, including atomic oxygen and CO. The vertical red line shows the cosmic standard abundance \citep{przybilla08} and the magenta line the latest solar abundance \citep{asplund05}, with the magenta arrow pointing toward the earlier value of \cite{grevesse98}. The blue line is the average UV-derived ISM abundance in the low-density warm gas that is least affected by depletion, with the blue arrow showing the corresponding value for higher mean density site lines \citep{cartledge04}. } \label{fig5}
\end{figure}

The average atomic oxygen abundance toward Sgr B2 derived here, $(2.51 \pm 0.69) \times 10^{-4}$ with respect to hydrogen nuclei, is in excellent agreement with the ISO value of $2.7 \times 10^{-4}$ \citep{lis01}, which was based on $^{13}$CO column density estimates for the molecular gas component. Figure~\ref{fig5} shows a normalized histogram of the O$^0$ abundances in the 120 individual velocity channels. The histogram is non-Gaussian and shows a narrow peak around $2.25 \times 10^{-4}$ and a broader shoulder around $3.15 \times 10^{-4}$, comparable to the values of $3.1 - 3.5 \times 10^{-4}$ derived by \cite{wiesemeyer16} toward W31C, G34.26, and W49N. The origin of the variations in the derived atomic oxygen abundance among different velocity components can be investigated further by using independent observations of additional molecular tracers, including the oxygen ions,  CH as a proxy for H$_2$ \citep{gerin10}, argonium as a proxy for purely atomic gas \citep{schilke14}, and ammonia as a tracer of high-density gas; such analysis is beyond the scope of the present paper. 

As a reference, the cosmic standard abundance of oxygen is $(5.75 \pm 0.4) \times 10^{-4}$, as measured in a representative sample of unevolved early B-type stars in nearby OB associations \citep{przybilla08}. The latest solar photospheric abundance is $4.57 \times 10^{-4}$ \citep{asplund05}, significantly lower than the earlier \cite{grevesse98} value of $6.76 \times 10^{-4}$. \cite{cartledge04} presented a comprehensive analysis of high-resolution \textit{Hubble} Space Telescope observations of O\,{\sc i} and H\,{\sc i} Ly~$\alpha$ UV absorption along 36 sight lines that probe a variety of Galactic disk environments. They derive an average O/H ratio of $3.90 \times 10^{-4}$ in the low-density warm gas that should be least affected by depletion. Sight lines of higher mean density are characterized by a lower average O/H ratio of $2.84 \times 10^{-4}$. Taking the higher value as a reference for the atomic oxygen abundance in the ISM gas, our average abundance of  $2.51 \times 10^{-4}$ on the line of sight toward Sgr B2 corresponds to about 35\% gas-phase oxygen depletion. CO will also contribute to the oxygen budget\footnote{Other oxygen-bearing gas-phase species contribute at a much lower level, e.g., the H$_2$O fractional abundance with respect to H$_2$ is in the range $3 - 7 \times 10^{-7}$ \citep{neufeld00,lis10}.} with a typical abundance of $1 \times 10^{-4}$ with respect to H$_2$ in the molecular gas, which is dominant at most velocities on this line of sight (Fig.~\ref{fig2}). Adding the two contributions, we derive an estimate of the total gas-phase O$^0$ + CO oxygen content toward Sgr B2 of $\sim 3 \times 10^{-4}$ with respect to hydrogen nuclei, about 25\% lower than the O/H value derived in the low-density warm gas from the UV measurements.

\begin{figure}
   \centering
   \includegraphics[trim=1cm 3cm 9cm 2cm, clip=true, width=0.99\columnwidth]{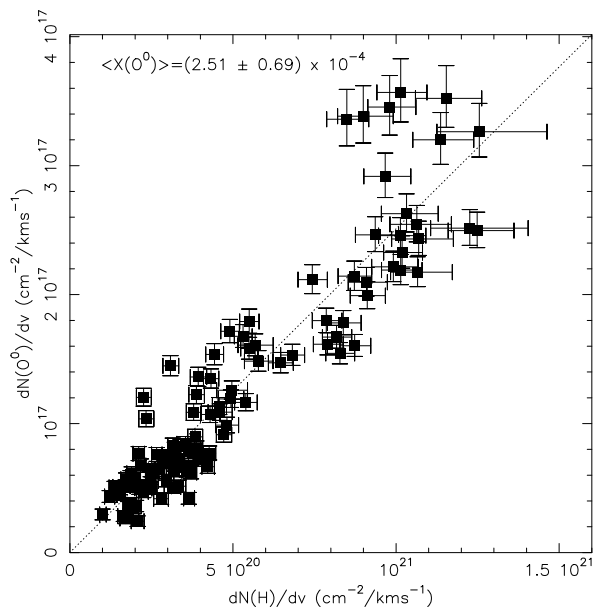}
   \caption{Atomic oxygen column density as a function of total hydrogen nucleus column density. Error bars are $1 \sigma$. The dotted line corresponds to a fractional abundance of $2.51 \times 10^{-4}$.} \label{fig4}
\end{figure}

Figure~\ref{fig4} shows the good correlation between the atomic oxygen and total hydrogen nucleus column densities (Pearson's correlation coefficient 0.85). The average abundance is $2.51 \times 10^{-4}$. Error bars mark the formal $1 \sigma$ uncertainties of the individual channels. We note that points with the highest O$^0$ column densities are located on average above the best-fit line. However, owing to the nonlinear dependence of the opacity on the line-to-continuum ratio at such high column densities, these points have large error bars and the result may not be significant. If we exclude points with the atomic oxygen column densities above $3 \times 10^{17}$ cm$^{-2}$, the resulting average atomic oxygen abundance is lower by only 2\%.

\section{Conclusions}
\label{sec:conclusions}
We have presented an analysis of archival SOFIA/GREAT observations of the [O\,{\sc i}] 63~$\mu$m absorption toward the Sagittarius B2(M) continuum source in the Galactic center. The high spectral resolution of the GREAT instrument allows, for the first time, the [O\,{\sc i}] absorption from individual line-of-sight clouds to be separated, velocity intervals affected by saturated absorption and telluric absorption to be masked, and accurate atomic oxygen column densities and abundances to be derived.
The atomic oxygen column density in the foreground spiral arm clouds toward Sgr B2 is well correlated with the total hydrogen column density, as determined from HF and H\,{\sc i} observations, with an average abundance of $(2.51 \pm 0.69) \times 10^{-4}$ with respect to H nuclei in individual 1~km\,s$^{-1}$ velocity channels. This value is in good agreement with the earlier ISO measurements on the same line of sight, and about 35\% lower than the average O/H ratio of $3.90 \times 10^{-4}$ in low-density warm gas derived from UV measurements \citep{cartledge04}. If we add a typical gas-phase CO content at $1 \times 10^{-4}$ with respect to H$_2$ in the molecular gas, which is dominant at most velocities on this line of sight (Fig.~\ref{fig2}), the total gas-phase oxygen content (O$^0$ + CO) on the line of sight toward Sgr B2 is $\sim 3 \times 10^{-4}$, or 300 ppm with respect to H, about 25\% lower than the low-density warm ISM oxygen abundance derived from UV measurements \citep{cartledge04}. With silicates and oxides contributing another 100 ppm \citep{whittet10}, the remaining oxygen fraction is about 175 ppm, which can be viewed as an upper limit for UDO on the line of sight toward Sagittarius B2. However, the expected ice contribution in this density regime is about 125 ppm \citep{whittet10}, leaving little room for UDO.

\begin{acknowledgements}
  
Based on observations made with the NASA/DLR Stratospheric Observatory for Infrared Astronomy (SOFIA). SOFIA is jointly operated by the Universities Space Research Association, Inc. (USRA), under NASA contract NAS2-97001, and the Deutsches SOFIA Institut (DSI) under DLR contract 50 OK 0901 to the University of Stuttgart. GREAT is a development by the MPI für Radioastronomie and the KOSMA/Universität zu Köln, in cooperation with the DLR Institut für Optische Sensorsysteme, financed by the participating institutes, by the German Aerospace Center (DLR) under grants 50 OK 1102, 1103 and 1104, and within the Collaborative Research Centre 956, funded by the Deutsche Forschungsgemeinschaft (DFG). Part of this research was carried out at the Jet Propulsion Laboratory, California Institute of Technology, under a contract with the National Aeronautics and Space Administration (80NM0018D0004). We thank B. Winkel for providing us with the H\,{\sc i} column densities used in the analysis and an anonymous referee for helpful comments regarding the overall oxygen budget. D. C. L. was supported by USRA through a grant for SOFIA Program 08–0038 (HyGAL).

\end{acknowledgements}

\begin{appendix}

\section{SOFIA data reduction}

  The data were collected on 2015 July 19, on the southern hemisphere deployment of SOFIA's Cycle 1, at 11.3 to 11.5~km altitude, under a precipitable water vapor column of typically 6~$\mu$m at zenith. The high-frequency channel of GREAT was tuned to the [O {\sc i}] line, alternating between the lower and upper sideband, so as to synthesize a sightline velocity interval from $-200$ to $+135$ km\,s$^{-1}$. In the velocity interval considered here, the median single-sideband system temperatures at zenith were 2100 and 2400 K for lower-sideband and upper-sideband tuning, respectively. Atmospheric and Galactic backgrounds were removed by chopping to a reference position at 160$^{\prime\prime}$ on both sides
of Sgr B2(M), at a position angle of $30^\circ$ (east to south), that is to say, perpendicular to the object's elongation. The FWHM beam size of 6.6$^{\prime\prime}$ was measured from cross-scans on Mars.

The spectra were calibrated to forward-beam brightness temperatures (with a 97\% forward efficiency) against loads at ambient and cold temperatures, and then to main-beam brightness temperatures using a 67\% main beam efficiency. The $63\,\mu$m [OI] transition is located in the wing of a broad water vapor absorption feature; the applied transmission correction was derived from modeling the measured atmospheric total power emission received in the signal and image bands. In order to minimize the impact of mixer gain drifts, only the off-target spectra immediately following a calibration load measurement were used (i.e., the correction was determined scan-wise and then scaled to the current elevation of each recorded spectrum).

The continuum emission of Sgr B2(M) was obtained from a dedicated double-sideband calibration. While the signal-to-image band gain ratio may deviate from unity (a standard deviation of 5\%), the overall reliability of the calibration scheme can be monitored with two tests: First, the saturation of the [OI] line at the systemic velocity defines the zero level for the single-sideband calibration. Second, the mesospheric [OI] line serves as a ``beacon'' that undergoes the same attenuation in the stratosphere as the astronomical signal. With these precautions, the calibrations and, consequently, the continuum levels of the lower- and upper-sideband tunings were brought into agreement. The spectra, $\mathbf{s}$, of Sgr B2(M) in the lower- and upper-sideband tuning are thus identical within the radiometric noise. The difference between the actually measured spectra ($\mathbf{y_\mathrm{L}}$ and $\mathbf{y_\mathrm{U}}$ for lower- and upper-sideband tuning, respectively) displays a well-defined linear baseline below $15\,\mathrm{km~s}^{-1}$. Instabilities that arise at velocities above this baseline are from the upper-sideband tuning and can be ignored there thanks to the redundancy with the lower-sideband tuning.
The spectra from the two tunings can then be expressed as
$\mathbf{y}_\mathrm{L} = \mathbf{s}+a_\mathrm{L}\mathbf{\nu},\,\, \mathbf{y}_\mathrm{U} = \mathbf{s}+a_\mathrm{U}\mathbf{\nu}\,,$
where $\nu$ is the frequency in the rest frame of Sgr B2(M). Thanks to the linear baseline fit to the difference spectrum, only two parameters (offset and slope) remained to be optimized, which was done in a way that ensures equal continuum levels on both sides
of the line-free portion of the spectra and reproduces the saturated absorption at systemic velocity. The good agreement between the two tunings in the overlapping velocity interval, from $-110$ to $+15\,\mathrm{km~s^{-1}}$, is taken as an assessment of the data processing algorithm.


\section{Atomic oxygen and HF column densities}

\begin{figure}
   \centering
   \includegraphics[trim=2cm 3cm 2.5cm 2cm, clip=true, width=0.99\columnwidth]{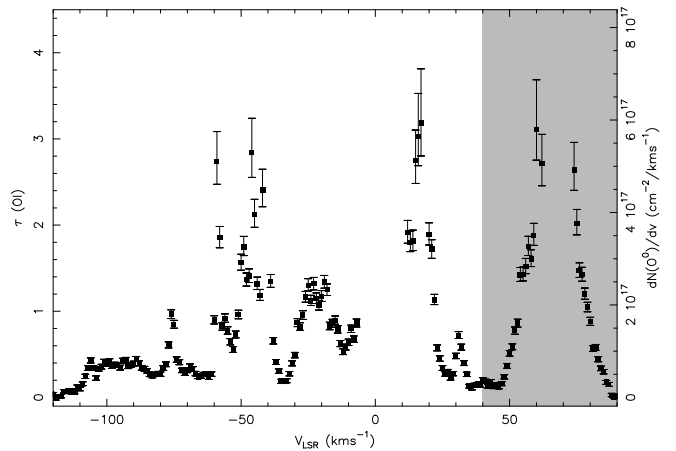}
   \caption{[O\,{\sc i}] optical depth (left axis) and atomic oxygen column density (right axis) in 1 km\,s$^{-1}$ channels as a function of velocity. Error bars are $1 \sigma$.  }
         \label{figb1}
\end{figure}

We converted the [O\,{\sc i}] optical depth to the atomic oxygen column density assuming that the absorbing gas covers the background continuum source and the entire population is in the lower state (e.g., \citealt{neufeld10}):

\begin{equation}
  \int \tau {\rm d} v  = { {A_{\rm ul} g_{\rm u} \lambda^3} \over {8 \pi g_{\rm l}} } \, N({\rm O^0}) = 5.365 \times 10^{-18} N({\rm O^0})\,\, {\rm cm^2\, km\, s^{-1}},
\end{equation}

\noindent where $A_{\rm ul}= 8.91 \time 10^{-5}\,{\rm s}^{-1}$ is the spontaneous radiative decay rate, $g_{\rm u} = 3$ and $g_{\rm l} = 5$ are the degeneracies of the upper and lower levels, and $\lambda = 63.184$ $\mu$m is the transition wavelength. Figure~\ref{figb1} shows the [O\,{\sc i}] optical depth (left vertical scale) and the resulting atomic oxygen column density (right vertical scale) in 1 km\,s$^{-1}$ velocity channels as a function of LSR velocity.  

\begin{figure}
   \centering
   \includegraphics[trim=2cm 3cm 2.5cm 2cm, clip=true, width=0.99\columnwidth]{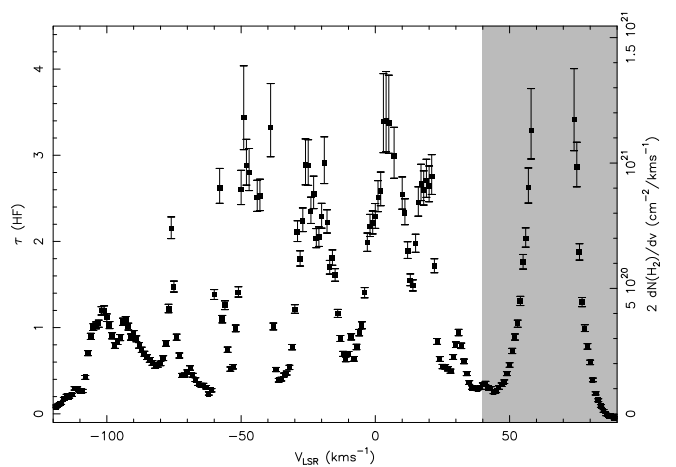}
   \caption{HF optical depth (left axis) and the H nucleus column density in the molecular component (right axis) in 1~km\,s$^{-1}$ channels as a function of velocity. Error bars are $1 \sigma$. }
         \label{figb2}
\end{figure}

\begin{figure}
   \centering
   \includegraphics[trim=2cm 3cm 4cm 2cm, clip=true, width=0.99\columnwidth]{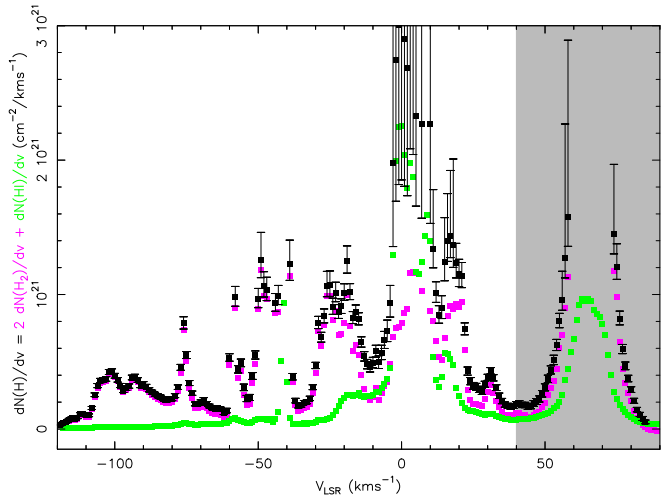}
   \caption{Total hydrogen nucleus column density toward Sgr B2(M) as a function of velocity (black squares). The molecular, $\rm 2\times N(H_2)$, and atomic, N(H\,{\sc i}), components are shown in cyan and green, respectively. Error bars are $\pm 1 \sigma$. An electronic table with the data used for this figure is available at the CDS.} \label{fig2}
\end{figure}

The corresponding formula for HF is\begin{equation}
  \int \tau {\rm d} v  = { {A_{\rm ul} g_{\rm u} \lambda^3} \over {8 \pi g_{\rm l}} } \, N({\rm HF}) = 4.157 \times 10^{-13} N({\rm HF})\,\, {\rm cm^2\, km\, s^{-1}},
\end{equation}
\noindent with $A_{\rm ul}= 2.42 \time 10^{-2}\,{\rm s}^{-1}$, $g_{\rm u} = 3$, $g_{\rm l} = 1$, and $\lambda = 243.2444$ $\mu$m. Figure~\ref{figb2} shows the HF optical depth and the resulting column density in 1 km\,s$^{-1}$ velocity channels as a function of LSR velocity. Figure~\ref{fig2} shows the total hydrogen nucleus column density as a function of velocity, along with the molecular and atomic contributions, as described in the text.

\end{appendix}

\end{document}